\documentclass{JHEP}

\def\a{\alpha}%GreekGreekGreek
\def\b{\beta}
\newcommand{\g}{\gamma}

\def\d{\delta}

\newcommand{\ep}{\epsilon}

\newcommand{\si}{\sigma}

\def\th{\theta}

\def\tr{{\rm tr}}
\def\ul#1{\underline{#1}}

\def\nt{{\tilde n}}
\def\Xt{{\tilde X}}

\newcommand{\non}{\nonumber\\}

\newcommand{\beq}{\begin{equation}}
\newcommand{\eeq}{\end{equation}}
\newcommand{\beqa}{\begin{eqnarray}}
\newcommand{\eeqa}{\end{eqnarray}}
\newcommand{\barr}{\begin{array}}
\newcommand{\earr}{\end{array}}
\newcommand{\ben}{\begin{enumerate}}
\newcommand{\een}{\end{enumerate}}
\newcommand{\bit}{\begin{itemize}}
\newcommand{\eit}{\end{itemize}}

\newcommand{\refeq}[1]{(\ref{#1})}

\def\vth{\vartheta}

\def\cF{{\cal F}}
\def\cR{{\cal R}}
\def \RR{{\cal R}}
\def\trr{\triangleright}

\title{
Matrix description of D-branes on 3-spheres~{}\footnote{Work supported 
by Polish State Committee for Scientific Research (KBN) under contract
5 P03B 150 20 (2001-2002)}
}

\author{      
Jacek Pawe\l czyk \\
 Institute of Theoretical Physics\\ Warsaw University, Ho\.{z}a 69,
PL-00-681 Warsaw, Poland\\[0.5ex] \email{Jacek.Pawelczyk@fuw.edu.pl}
}
\author{Harold Steinacker \\ 
 Laboratoire de Physique Th\'eorique et Hautes Energies\\
        Universit\'e de Paris-Sud, B\^atiment 211, F-91405 Orsay \\
and \\
Sektion Physik der Ludwig--Maximilians--Universit\"at M\"unchen\\
Theresienstr.\ 37, D-80333 M\"unchen  \\[0.5ex]
\email{Harold.Steinacker@physik.uni-muenchen.de}
}

\abstract{We discuss a  matrix model for D0-branes on $S^3\times M^7$  
based on quantum group symmetries. 
For finite radius of $S^3$ (i.e. for finite $k$), 
it gives results beyond the reach of  the ordinary matrix model. 
For large $k$ all known static properties of branes on  $S^3$ are reproduced.
}
\preprint{LMU-TPW 08/01 \\LPT-ORSAY 01-66 \\
  IFT-01/23}

\begin{document}

\section{Introduction}

Recently it has been discovered that in a special limit the dynamics of
D-branes can be described by a finite matrix  model \cite{myers}. 
Together with works on NCG in string theory \cite{ncg}
this has
sparked new interest in   strings propagating in  antisymmetric tensor fields
backgrounds. We are interested here in 
the dynamics of  D0-branes in a background of the form
$S^3\times M^7$. In the approximation of 
infinite radius of $S^3$, i.e. infinite $k$, the dynamics is  
given by the Myers action \cite{myers}. 
This action can be applied in various situations and 
leads to numerous interesting effects (see e.g. \cite{other}).
Unfortunately, it can not describe the finite $k$ effects known e.g. from the 
DBI action \cite{jbds} or the WZW description of strings in NSNS 
background \cite{as,ars,ars2}.

In this paper we would like to go beyond this approximation. 
We discuss the dynamics of D0-branes on $S^3\times M^7$ spaces with a NSNS 
antisymmetric tensor field background for
finite radius $k$ of $S^3$. Our aim is to 
construct a matrix model which will give results as good as the DBI action for 
Dp-branes,  and which 
might be a building block of a full D0-brane theory. Our approach is
inspired by \cite{fuzzyqsphere}.
The actions will be constructed by imposing a certain symmetry, 
which is a version of
the quantum group (more precisely quantum universal enveloping algebra)
$U_q(so(4))$. The name  q--matrix
model originates from this algebra. 
Although we do not propose any definite form of the action, the 
$U_q(so(4))$ symmetry is restrictive enough so that we can
discuss several physical properties of the model. 
For example, we can given an
expression for the energy of the wrapping which is given by the 
quantum dimension
of a representation, and we can solve equations of
motion and obtain quantum 2--spheres. 

The outline of the paper is the following.
In the rest of this introduction we shall recall some facts concerning the
macroscopic description of branes in terms of the DBI action and 
the microscopic
theory of many D0-branes of Myers \cite{myers}. In the latter we emphasize the 
approach of \cite{ars,ars2} which uses an 
$SU(2)$ WZW model to derive the D0-branes 
action. In Section 2 we shall present the $U_q(so(4))$ algebra, 
as well as some modules and invariants. It
will play the crucial role in constructing the matrix model. Then we shall
define the model and discuss its predictions.
Section 3 is devoted conclusions and the  discussion of  problems and 
limitations of the model. 
Appendices contain our conventions, some results on
representations of $U_q(su(2))$ and two proofs.

{\bf DBI branes in NSNS background.}
The  string theory background  of interest here  has the form
$S^3\times M^7$ with some nonzero $H^{NSNS}$ and constant dilaton. The most
studied example is  obtained as the near horizon limit of the  
F1, NS5 system (see e.g. \cite{mald-strom}). 
Below we write only the relevant
fields living on $S^3$
\beqa
ds^2/\a' =k\; d\Omega_3^2,\quad
H^{NSNS}/\a'= 2k\;\ep_3,\quad
e^{2\phi } = const.\label{back}
\eeqa
where $k$ denotes the number of NS5-branes and  equals to the level
of 
the appropriate $\widehat{su(2)}_k$ WZW model (see below), and 
$\ep_3$ is the volume element of the 
unit 3-sphere. 

The action of the D-branes is given by the DBI expression  
which properly describes the dynamics of branes for large (but finite) $k$.
In \cite{jbds} classical configurations of
Dp-branes  embedded in \refeq{back} were considered. The branes were embedded
in such a way that two of the dimensions of the brane
wrapped $S^2\subset S^3$ and the rest extended in $M^7$ forming an effective
D(p-2)-brane. 
It appears that the position $\vth$ of the spherical brane on $S^3$ is
quantized
\beq\label{fquant}
\vth\,=\frac{\pi n}k,\qquad 0\leq n \leq k
\eeq
where  $\th$ is the standard polar angle.
The effective D(p-2)-brane has a
tension which takes into account the energy of the wrapping. The
expression for this tension is 
$T_{(p-2)}^{eff}=T_p\, 4\pi\a'\;  k \sin\frac{n\pi}k$.
One should expect that $n=0$ corresponds to a single effective D(p-2)-brane,
which contradicts the above formula. Luckily, due to the curvature terms
this result gets corrected by shifting $n\to n+1$ \cite{as-d2},  
\beq\label{energy}
T_{(p-2)}^{eff}=T_p\, 4\pi\a'\;  k \sin\frac{(n+1)\pi}k
\eeq
Now for large
$k$ and $n=0$ one gets the proper relation $T_{(p-2)}^{eff}=T_p\, 4\pi^2\a'$.

We should add that the very similar results were also obtained  
for the RR background in \cite{j-rey}. 

{\bf Matrix model and ${\widehat\mathbf{su(2)_k}}$ WZW model.}
The above system can also be studied in the $1/k$ expansion 
as a matrix model of many D0-branes \cite{myers}. The leading
terms of the appropriate action is obtained approximating the $S^3$ by a
3-plane and taking 
$B=\frac13 k \a' \ep_{ijk}X^i dX^j\wedge dX^k$.
Then 
\beq\label{m-action}
S\sim T_0 \tr\left[1-\frac{(2\pi^2)}{k^2}
\;(\;\frac14 \,[X^i,X^j]^2+\frac i3\ep_{ijl}X^i[X^j,X^l]\;)+\dots\right]
\eeq 
The equations of motion for $X$ are solved by 
\beq\label{fuzzys}
[X^i,X^j]=i\ep^{ijk}X^k
\eeq
where now $X$'s are $N\times N$ matrices of positions of  $N$ D0-branes.
For the fixed central element of the above algebra (the second Casimir) this
defines the so-called fuzzy sphere \cite{johnm}. It appears that the fuzzy spheres
have the dynamics and quantum numbers as ordinary D2-branes described in the
previous subsection \cite{ars2,jbds}. 
One can rederive  \refeq{fuzzys} using CFT language \cite{ars2}. 
We have decided to
recall some facts from this approach as it 
directly leads to the q-deformed case.

The analysis of the $\widehat{su(2)}_k$ WZW model with the simplest boundary 
conditions\footnote{The boundary conditions break current algebra
 $su_L(2)\times su_R(2)\to su_V(2)$. This should  be related to the form of the 
  twisting of $U_q(so(4))$ we have chosen in Sec.\ref{sec:so4} but the precise 
  connection is not clear to us.}\label{foot:break}
defining branes shows that D--branes  are classified 
by an integer $\nt$ which satisfies 
$0 \leq \nt \leq k$. The Hilbert space of the boundary 
CFT decomposes into irreducible representations 
of the affine Lie algebra $\widehat{su(2)}_k$. The ground ring of the
associated boundary operators $Y^I_a(x)$, 
($I$ ranges from $0$ to $\min (\nt,k-\nt)$, and $-I \leq a \leq I$) form
irreducible  spin $I$ representations of the horizontal $su(2)$ algebra.
Their OPE's are  given by \cite{ars}
\beq\label{b-algebra}
Y^I_a(x) Y^J_b(y) = 
\sum_{K,c} (x-y)^{h_K-h_I-h_J}
\left[\begin{array}{ccc}I&J&K\\ a&b&c \end{array} \right]
\left\{ \begin{array}{ccc} I&J&K\\ \nt/2& \nt/2 & \nt/2 \end{array} \right\}_q 
             Y^K_c(y)+\dots,
\eeq
where $q=e^{\frac{\pi i}{k+2}}$, the sum goes from 
$K=0$ to
$\min(I+J,k-I-J,\nt,k-\nt)$, $-K\leq c \leq K$ 
$h_I=I(I+1)/(k+2)$ is the  conformal weight of $Y^I_a(x)$,
the first bracket denotes the Clebsch--Gordon coefficients of $su(2)$
and the curly brackets denote the $q$--deformed $6J$--symbols of
$U_q(su(2))$.

{\bf $\mathbf{k\to\infty}$ limit.} It is clear that  in $k\to\infty$ limit,
$h_I\to 0$ and the $q$--deformed $6J$--symbols tend 
to the ordinary $6J$--symbols so that the  OPE becomes the 
associative matrix algebra  $Mat(\nt+1)=(\ul1\oplus\ul3\oplus
\dots \oplus (2\nt+1))$. From the point of view of
strings moving in the other 7 dimensions $Mat(\nt+1)$  is a kind of
Chan-Patton algebra. It carries 
information about the internal geometry of the brane. In fact, $Y^I_a$ can be
thought of as spherical harmonics with an unusual 
multiplication law.
It is known as the algebra of functions of the fuzzy sphere \cite{johnm}.
The first descendants (gauge fields) $j_{-1}^i Y^I_a$ of the ground ring   
have 
conformal dimensions $h_I^{(1)}=1$ in this limit. Thus
they form vertex operators describing the dynamics of the wrapped D-branes. 
These were used in \cite{ars2} in order to find the
effective action \refeq{m-action} for the $X^i$ matrices.

{\bf Finite k.} Now  $Y^I_a$ do not form any algebra, as 
$h_I$ are no longer zero\footnote{
Some of the
operators can have large dimensions: for maximal $I\sim k/2$ this can be
about  $h\sim k/4$.}.
If one naively drops factors $(x-y)^{h_I+h_J-h_S}$,
then  $Y^I_a$ form a quasiassociative algebra, which is not a 
very convenient structure to work with. 

It was noticed in \cite{ars,fuzzyqsphere} that quantum groups come to rescue: 
one can recover the associativity of 
the algebra at the price of twisting it. Technically this means that one
changes the ordinary Clebsch--Gordon coefficients in \refeq{fuzzys} 
by their q-deformed
versions. Then $Y^I_a$ form representations of $U_q(su(2))$ 
(which will be referred to as $U_q(su(2))^V$ below). 
For $\nt\in (0,[k/4])$, the tensor product of any two elements
$Y^I_a$ is completely reducible, and after this twisting one 
obtains the associative algebra 
$Mat(\nt+1)=(\ul1\oplus\ul3\oplus \dots \oplus (2\nt+1))$. 
For $\nt > [k/4]$ this is no longer
the case\footnote{For $\nt > [k/4]$, quasiassociativity
cannot be removed completely due to the truncation in the above algebra. We
will not worry here about the precise construction of fields in general.}; 
however our results below 
continue to make sense even for $\nt > [k/4]$, and will 
be manifestly invariant under $\nt\to k-\nt$. 
The position matrices (analogs of $X^i$) 
correspond again to the gauge field operators $j_{-1}^i Y^I_a$. 
They are also vectors in  $\ul3$  of $U_q(su(2))^V$. 

Below we shall significantly extend this approach and analyze its physical
content.

\section{q-matrix model}

This section is devoted to the construction and the analysis of a 
matrix model 
based on a certain q-deformed symmetry algebra.
Our basic variables are position matrices 
$M^\mu$ $(\mu=1,\dots 4)$
describing a ``quantum 3-sphere'' $S^3_q$ and  
representing positions of a system of $N=\nt+1$ objects which we shall call
q-D0-branes\footnote{We add the prefix "q-" here because it might be that 
these objects are different than the usual D0-branes.}. The physical
interpretation of these fields is similar to $X^i$'s of
\refeq{m-action}. Under the "internal" 
$U_q(su(2))^V$ they shall transform 
in the quantum adjoint action
\beq
U_q(su(2))^V\ni u\trr M^\mu=u_1 M^\mu S(u_2),
\eeq
where $u_{1,2}$ are $\nt+1$ dimensional matrices
obtained by taking the representation 
of the coproduct $\Delta(u) = u_1 \otimes u_2$.
The position matrices 
$M^\mu$  belong also to the \ul4-dimensional module of the space-time
symmetry 
algebra which will be a twisted version of $U_q(so(4))$
denoted by $U_q(so(4))_\cF$: this will be an isometry of 
the ``quantum 3-sphere'' $S^3_q$. 
We shall also  assume that classical single brane configuration
will break   
$U_q(so(4))_\cF$ to $U_q(su(2))^V$ 
of the previous section (see footnote \ref{foot:break}). 
Technically this
means that $U_q(su(2))^V$ should be Hopf subalgebra of $U_q(so(4))_\cF$.
This will be crucial for the predictive power of the model. 
From now on, all group-theoretic objects (invariant tensors etc.)
are understood to carry a label $q$, which is suppressed.
For the notations we refer to Appendix \ref{app:tens} and \cite{fuzzyqsphere}.

Before we go into the details we should make a
comment. 
One must be aware that $q=e^{\frac{\pi i}{k+2}}$ involves $k$,
which is inverse proportional to the curvature of the $S^3$. This means that
the process of 
changing symmetry algebras to their q-deformed versions takes into account the
so-called sigma model corrections of the string theory, which are small for
large $k$. On the other hand, our
analysis is based only on the $S^3$ part of the model, and
it seems that one can not hope to get the full answers for finite $k$. 
Hence the situation is unclear at the moment, and 
the results obtained below should be treated with some caution.   
In the end of this section we shall compare the 
q-matrix model results with some  results obtained by other means 
for large $k$ and show their astonishing agreement.

\subsection{$\mathbf{U_q(so(4))}$}
\label{sec:so4}

{\bf Algebra.} The aim of this section is to construct a 
suitable quantum ``group'' (Hopf algebra) which describes the symmetries
of a quantum 3 sphere $S^3_q$, formally defined 
by \refeq{q3sphere}, 
and
which is compatible with a ``vector'' $U_q(su(2))^V$ symmetry. 
We expect that this isometry algebra 
should be a version of the quantum group $U_q(so(4))$.
As an algebra, the ``standard'' $U_q(so(4))$ is simply the tensor product
of two commuting $U_q(su(2))$ algebras, 
$U_q(so(4)) = U_q(su(2))^L \otimes U_q(su(2))^R$. 
It carries naturally
the structure of a (quasitriangular) Hopf algebra.
There is a natural embedding of the ``vector'' (sub)algebra 
$U_q(su(2))^V\rightarrow U_q^L
\otimes U_q^R$:  
\beqa
d: u \mapsto (u_1 \otimes u_2) = \Delta(u)
\label{vector-alg}
\eeqa
where $\Delta(u)$ is the coproduct of $U_q(su(2))$.
The problem with this definition is that the embedding $d$ is not 
compatible with the standard coproduct of $U_q(su(2))^L \otimes
U_q(su(2))^R$. This means that $U_q(su(2))^V$ will not act 
correctly on products of fields (i.e. matrices, here).

The solution to this problem is provided by twisting. For the general
theory of twisting    
we refer to \cite{ch-p}, and to \cite{majid} (Section 2.3).
Consider the modified coproduct
\beqa
\Delta_{\cF}: U_q^L \otimes U_q^R &\rightarrow&
    (U_q^L \otimes U_q^R) \otimes 
     (U_q^L \otimes U_q^R), \non
   u^L \otimes u^R &\mapsto& 
    {\cF}(u^L_1 \otimes u^R_1) \otimes (u^L_2 \otimes u^R_2) {\cF}^{-1}
\label{coprod-so(4)-new}
\eeqa
where ${\cF} = \cR{}_{32} \in U_q \otimes U_q$
is the universal $R$--''matrix'' of $U_q(su(2))$ with reversed
components.
The subscripts of $\cR$ refer to the positions in the tensor product.
This  defines again a quantum group, i.e. a quasitriangular Hopf 
algebra.   
We will denote this (twisted) quantum group with
$U_q(so(4))_{\cF}$. It is easy to check that it satisfies 
\beq
\Delta_{\cF} \circ d = (d \otimes d) \circ \Delta,
\eeq
i.e. $U_q(so(4))_{\cF}$ is
compatible with the embedding \refeq{vector-alg}
of the quantum group $U_q(su(2))^V$.
This is what we were looking for. 
In fact if $q$ were real, this $U_q(so(4))_{\cF}$ would coincide precisely 
with the $q$-Lorentz group $U_q(so(3,1))$ \cite{wessetal}, 
and much is known about its algebraic structure and representations.
However since we are looking for matrix models in a 
rather different setting, we shall construct the modules of interest 
(matrices) and the corresponding invariants 
using an ad--hoc approach, trying to minimize the technical background. 

{\bf Modules.} 
Here we discuss the $\ul4$ module  
of $U_q(so(4))_{\cF}$ to which the position matrices $M^\mu$ belong. 
There are two obvious ways to construct it: consider $2 \times 2$ matrices 
$X = X^{\dot{\a}}_{\;\;\b}$ and $\Xt = \Xt^{\a}_{\;\;\dot{\b}}$, where
dotted indices refer to $U_q^L$ and undotted indices to 
$U_q^R$. The naive guess would be that they transform as 
$X \rightarrow u^L X S(u^R)$ and $\Xt \rightarrow u^R \Xt S(u^L)$.
However, this is not compatible with the requirement that 
products of matrices transform using the coproduct of $U_q(so(4))_{\cF}$,
i.e. that these matrices form a  $U_q(so(4))_{\cF}$--module algebra. 
However, the following action of $U_q(so(4))_{\cF}$ 
defines a consistent $U_q(so(4))_{\cF}$--module algebra:
\beq
(X,\Xt) \mapsto u \trr_{\cF} (X,\Xt)   
  = (u^L X S(u^R),
      \cR{}_2 u^R \cR{}_b^{-1} \Xt S(\cR{}_1 u^L \cR{}_a^{-1}))
\label{M-trafo-new}
\eeq
extended to products via the new coproduct \refeq{coprod-so(4)-new}. One
verifies that products transform as 
\beq
\label{M-traf-l}
X\Xt \rightarrow u^L_1 X \Xt S(u^L_2), \quad \Xt X \rightarrow u^R_1 \Xt X S(u^R_2),
\eeq
see Appendix \ref{app:cop}.
The vector subalgebra $U_q(su(2))^V$ then acts as 
\beq
u^V \trr_{\cF} (X,\Xt) = (u^V_1 X S(u^V_2), u^V_1 \Xt S(u^V_2)),
\label{M-vector}
\eeq
as it should. Also,
$X\Xt \rightarrow u^V_1 X \Xt S(u^V_2)$ and 
$\Xt X \rightarrow u^V_1 \Xt X  S(u^V_2)$. This shows that all the objects (algebras
and modules) have the required properties.

Since there is only one representation $\ul4$ 
of $U_q(so(4))_{\cF}$, there should be a relation between these
two actions on $X$ and $\Xt$. Indeed, one can identify  $X$ and $\Xt$  via 
\beq
\Xt = -\cR{}_2 q^H S(X) S(\cR{}_1).
\label{Y-constraint}
\eeq
Here $S$ is the induced antipode of $U_q(su(2))$ on the $2 \times  2$ 
matrix $X$.
It is easy to verify that this indeed induces  the action \refeq{M-trafo-new}
on $\Xt$. This restricts the degrees of freedom as required, and 
will allow us to define a suitable quantum 3-sphere.
We use this to construct a basis for $\Xt$ modules in terms of the
basis of $X$ modules. Let us call the latter 
$\si_\mu=(iq,\si_i)$. 
Then the former called
${\tilde \si}_\mu$ are given by
${\tilde \si}_\mu=(-iq^{-1/2},q^{1/2}\si_i)$ (see Appendix \ref{app:proof}).

{\bf Invariants.} Using these results, one sees immediately that 
$q$-traces ($\tr_q(A)\equiv \tr(Aq^{-H})$)
\beqa
S^{(n)}_1= \tr_q ((X\Xt)^n) \qquad
S^{(n)}_2 = \tr_q ((\Xt X)^n), \quad n\in \mathbf{ N}
\label{terms}
\eeqa
are invariant under $U_q(so(4))_{\cF}$. 
Explicitly, in  case of interest ($2 \times 2$ matrices $X,\, \Xt$) 
we have $q^{-H} ={\rm diag}(q, q^{-1})$.
For example, the invariant tensor for 
the representation $\ul4$ is obtained by
$q^{1/2}[2]\; g_{\mu\nu}=\tr_q(\si_\mu{\tilde \si}_\nu)$. 
It has the block diagonal form
$g_{\mu \nu} =$
$\left(\begin{array}{cc} 1 & 0 \\ 0 & g_{ij} \end{array} \right)$
where $g_{ij}$ is the $U_q(su(2))$
invariant metric defined as above from $\si_i$ 
matrices and given explicitly in Appendix \ref{app:tens}.

\subsection{The model and its comparison with known results} 

\def\Mt{{\widetilde M}}
\def\vM{{\vec M}}

We would like to construct a ``quantum'' 3-sphere $S^3_q$ 
which will play the 
role of the $S^3$ from the string geometry \refeq{back}. Thus we
consider matrix valued $Mat(\nt+1)$ fields $M^\mu$
in $\ul4$ of $U_q(so(4))_\cF$ living on a  $S^3_q$.
Using those variables we shall construct
candidates for actions describing the dynamics of (quantum) D0-branes.

The techniques of the previous subsection give a convenient way of creating
$U_q(so(4))_\cF$--invariant tensors, with the help of \refeq{terms}. 
Let us introduce 
\beq 
M=\si_\mu M^\mu,\quad \Mt={\tilde \si}_\mu M^\mu.
\eeq
Then the formal ``quantum``
3-sphere $S^3_q$ is defined by the relation 
$\tr_q(M\Mt)=q^{1/2}[2]R^2$, i.e. 
\beq\label{q3sphere}
(M^4)^2+g_{ij}M^jM^j=R^2\cdot 1,\quad R>0.
\eeq  
It is natural to expect
that the radius of this\footnote{We
shall not discuss its  precise algebraic/geometric meaning.}  
sphere is $R^2\sim k$ in analogy with the string
background \refeq{back}. For the future application we shall often set 
$R^2=\a^2 k$.

We assume that the action of the q-matrix model can be expressed as power
series in $M$ and $\Mt$ (see \refeq{terms}). Due to the above constraint 
we have
\beqa\label{mmt}
M\Mt =q^{1/2}[R^2+F^l_L \si_l],\quad \Mt M=q^{1/2}[R^2+F^l_R \si_l]
\eeqa
where
\beqa
F^l_L&=&i( q M^4M^l-q^{-1}M^lM^4)-\ep^l_{ij}M^iM^j\\
F^l_R&=&i(-q^{-1}M^4M^l+ q M^lM^4)-\ep^l_{ij}M^iM^j
\eeqa 
are vectors of  $U_q(su(2))^L$ and $U_q(su(2))^R$, respectively.
We call them L (R) field-strength as, in the limit $q=1$, they are equal 
and correspond to the
ordinary field strength of \cite{ars2}. But notice that we can not write any
Chern-Simons term here.
From \refeq{mmt} it follows 
that any polynomial in $M$ and $\Mt$ can be also written in terms of
$F^l_{L,R}$. The examples are $g_{ij}F^i_LF^j_L,\; g_{ij}F^i_RF^j_R,\;
\ep^n_{ij}F_{Ln}F^i_LF^j_L$, where $\ep$ is an invariant 3-index tensors of
$U_q(su(2))^{L,R}\subset 
U_q(so(4))_\cF$ (see Appendix \ref{app:tens}).

Now we can easily construct the invariant actions. One must remember that
$F^l_{L,R}\in Mat(\nt+1)$ i.e. they transforms under $U_q(su(2))^V$. 
Explicitly\footnote{This can be reconciled with the concept of a differential 
calculus using the concept of frames, see \cite{fuzzyqsphere}.
}
\beq
U_q(su(2))^V\ni u\trr F^l_{L,R}=u_1 F^l_{L,R} S(u_2).
\eeq 
The  invariants 
are expressed as q-traces $\tr(Aq^{-H})$. 
Now the trace is over the ``internal'' degrees of freedom, i.e. over
$Mat(\nt+1)$.
The matrix $H$ is 
determined by the  representation we want to work with.
This is a somewhat
unusual and unpleasant property (compared to the standard matrix model) of the
model, but at the moment we do not know  how to overcome it.
The representations can be reducible (describing sets of
clusters of q-D0-branes) or irreducible. 
In the latter case it has the following 
form $H={\rm diag}(\nt,\nt-2,\dots ,-\nt)$.

The simplest terms which could contribute to
the action are of the from
\beq\label{actions}\tr_q(1),\quad 
\tr_q(g_{ij}F^i_LF^j_L),\quad \tr_q(g_{ij}F^i_RF^j_R),
\quad\tr_q(\ep^n_{ij}F_{Ln}F^i_LF^j_L),\;\dots 
\eeq
We shall not try to find out
the precise form of the relevant action. This would require a detailed
analysis of the 
 string theory on $S^3\times M^7$ which, to our knowledge, is not available
 now. Thus we shall limit the discussion to the most basic properties of 
{\it any action} which can be formed in our case.  

First of all we  notice that 
\beq\label{eqm}
F^l_L=F^l_R=0
\eeq
solves  any possible equations of motion. Moreover \refeq{eqm}
nullifies all of the terms \refeq{actions} except $\tr_q(1)$.  The
energy of the classical configuration which corresponds to the irreducible 
representation is therefore given by so-called quantum dimension
\beq\label{qdim}
\tr_q(1)=[\nt+1]_q=\frac{\sin(\frac{(\nt+1)\pi}{k+2})}{\sin{(\frac\pi{k+2})}}.
\eeq
For $k>>1$ this gives (up to an overall constant) the same function of $k$ as
appearing  in the tension of the effective D(p-2)-brane 
\refeq{energy}. The proper identification of $n$ in \refeq{energy} and $\nt$ in
\refeq{qdim} gives $\nt=n$. The formula has
all expected features of the energy of $N=\nt+1$ D0-branes on $S^3$. It might be
that this is an exact result at least at genus zero level.

The equations of motion  \refeq{eqm} can be rewritten in
more explicit form\footnote{algebraically, these are the same as the relations 
of a ``light-cone'' in quantum Minkowski space \cite{wessetal}.}
\beqa
[M^4, M^l] &=& 0, \non
\label{eq-m}
\ep_{ij}^l \; M^i M^j &=& i (q-q^{-1}) M^4 M^l 
\eeqa
Due to $[M^4, M^l]=0$ we can diagonalize $M^4$ (hence also  $g_{ij}M^iM^j$). 
The algebra generated by 
solutions of \refeq{eq-m} of the form $\{M^4\sim \mathbf{1},M^i\}$  
is called the q-fuzzy sphere $S^2_{q,\nt}$ \cite{fuzzyqsphere}.
This is a slight generalization of a Podles sphere \cite{podles},
$q$ being a phase. It is characterized by the radius $r_\nt$ of the sphere
defined by $r_\nt^2=g_{ij}M^iM^j$. One finds
\beq\label{radius}
r_\nt^2=\ R^2 \left(1-\frac{[2]_{q^{\nt+1}}^2}{[2]_q^2}\right) = 
       \ R^2 \left(1- \frac{\cos^2(\frac{(\nt+1)\pi}{k+2})}
                           {\cos^2(\frac{\pi}{k+2})}\right).
\eeq
Notice that this formula is $\nt\to k-\nt$ invariant. The corresponding
4--component is naturally given by 
$M^4 = R\; {[2]_{q^{\nt+1}}}/{[2]_q}$ 
$= R\; {\cos(\frac{(\nt+1)\pi}{k+2})}/{\cos(\frac{\pi}{k+2})}$,
which covers positive and negative values.
For large $k$ and $1\ll\nt\ll k$, \refeq{radius} is well approximated by 
$r_\nt^2\approx R^2 \sin^2( \frac{\nt\pi}{k})$ which is the classical expression for
the radius of $S^2\subset S^3$
 as follows from \refeq{fquant} for $R^2=k$. Again we get  astonishing
agreement although
the formula \refeq{radius}  has completely
different origin, namely the representation theory of quantum algebras.

Let
us also make the approximation of small spheres $S^2_{q,\nt}$  
and substitute 
$M^4=\sqrt{k}+O(g_{ij}M^iM^j)$ (we set $R^2=k$ here). Then 
$x^l=i{M^l}/({(q-q^{-1})\,\sqrt{k}})$ respects 
q-fuzzy sphere equation as written in \cite{fuzzyqsphere},
\beq\label{fuzzys2}
\ep_{ij}^l \; x^i x^j=x^l
\eeq
In the large $k$ approximation  the above becomes the ordinary fuzzy sphere
\cite{johnm} as obtained in \cite{ars2}. 

We have also checked the spectra of the fluctuations and they slightly differ
from what 
is known for the WZW model \cite{jbds} and the DBI action \cite{j-rey}. 
However the comparison we have just made might be too naive, because in
order  
to find masses we need also the time component of the dynamics which is 
beyond the scope of this work.

\section{Problems and conclusions}

We conclude this paper with some comments concerning the obtained results. It
seems astonishing that such a simple model based on almost 
no physical input
precisely reproduces {\em static} properties of the D0-brane system. 
The origin of
this success lies in the applied symmetries.  
One can see that formula \refeq{qdim} follows from 
the assumption that $Mat(\nt+1)$ is in a representation of $U_q(su(2))^V$,
but it is also 
intimately related to the fact that $U_q(so(4))_\cF$ constrains the simplest
equations of motion $F^i_{L,R}=0$ in such a way that the other contributions 
to the brane energy vanishes. 
Next, the nice equations of motion \refeq{eq-m} directly follow from the twist
$\cF$ in $U_q(so(4))_\cF$: the latter has been imposed on physical grounds,
i.e. by requiring the unbroken $U_q(su(2))^V$ symmetry.
The equations of motion yield an attractive formula for the 
radius of  $S^2_{q,\nt}$. {\it Last but not the least}, we must stress that all
obtained formulae work for all boundary conditions $\nt\in [0,k]$ suggesting 
that $S^2_{q,\nt}$ are well defined even for these $\nt$ which exceeds the
limit of ``nice''(irreducible highest weight) representations of $U_q(su(2))$.

One should expect that the model has some limits of applicability. 
It works for large $k$, but the symmetry alone is not able to pick any specific
action. It seems inevitable that in order to achieve more 
accurate results one must 
understand also the dynamics of $M^7$ degrees of freedom.
Besides one must
realize that the model does not have the original isometry of $S^3$. We do not 
believe that the isometry is broken by the quantum effects (due to
$q=e^{i\pi/(k+2)}$) discussed in the previous section. 
Hence  in the final version of the model, there should be the
possibility to twist back the symmetry algebra from $U_q(so(4))_\cF$ to
$U(so(4))$. Furthermore, the model requires to fix the representation of
$U_q(su(2))^V \subset U_q(so(4))_\cF$ in order to define q-traces. This sounds
bizarre and awaits further clarification.
Next, the model has problems with reality conditions for the $M$
matrices. More precisely, we were not able to find any reality conditions which
would be compatible with the algebra actions. 
Finally the gauge symmetry U$(\nt+1)$ of the standard matrix models is obscured here.

In spite of all those unsettled issues the model reproduces all the static properties
of the D0-brane system, with astonishing accuracy.
These results are beyond reach of the standard matrix model.  The key to the
success lies in the  quantum symmetry. Its  
relevance  to string theory clearly deserves further studies.

\acknowledgments{
The authors thank J.Lukierski and the organizers of the XXXVII Winter School
of Theoretical Physics in Karpacz for hospitality and stimulating atmosphere
during the school. 
We would like to thank Chong-Sun Chu, J.Kowalski-Glikman, 
H.Grosse, B.Jurco, J.Madore, and W. Zakrzewski  
for inspiring discussions. 
We are also grateful for
hospitality in the Laboratoire de Physique Th\'eorique in Orsay and
Sektion Physik der Ludwig--Maximilians--Universit\"at in M\"unchen.
H.S. thanks the DFG for a fellowship.

}

%%%%%%%%%%%%%%%%%%%%%%%%%%%%app.tex%%%%%%%%%%%%%%%%%%%%%%%%%
%\newpage
\begin{appendix}

\section{Appendices}

\subsection{Basic properties of $U_q(su(2))$}

The basic relations of the Hopf algebra $U_q(su(2))$ are 
\beq\label{U-q-rel}
[H, X^{\pm}] = \pm 2 X^{\pm},  \qquad
[X^+, X^-]   =  \frac{q^{H}-q^{-H}}{q-q^{-1}}=  [H]_{q},
\eeq
where the $q$--numbers are defined as 
$[n]_q = \frac {q^n-q^{-n}}{q-q^{-1}}$. 
The action of $U_q(su(2))$ on a tensor product of representations 
is encoded in the coproduct
\beqa 
\Delta(H)       &=& H \otimes 1 + 1 \otimes H \nonumber \\
\Delta(X^{\pm}) &=&  X^{\pm} \otimes q^{-H/2} + q^{H/2}\otimes X^{\pm}.
\label{coproduct-X} 
\eeqa
We will use the Sweedler--notation $\Delta(u) = u_1 \otimes u_2$, where a 
summation convention is understood.
The antipode and the counit are given by
\beqa
S(H)    &=& -H, \quad  
S(X^+)  = -q^{-1} X^+, \quad S(X^-)  = -q X^-, \nonumber \\
\epsilon(H) &=& \epsilon(X^{\pm})=0.
\eeqa
Moreover, the following element  
\beq
v = (S\RR_2) \RR_1 q^{H}
\label{v}
\eeq
is in the center of $U_q(su(2))$, and has the eigenvalues
$q^{2j(j+1)}$ on the spin $j$ representation.

\subsection{Representations and invariant tensors}\label{app:tens}

The  $q$--deformed sigma--matrices, i.e. the Clebsch--Gordon coefficients
for $(3) \subset (2) \otimes(2)$, are given by
\beq
\sigma^{+ +}_{1} = \sqrt{[2]_q} = \;\sigma^{- -}_{-1}, \qquad
\sigma^{+ -}_{0} = q^{\frac12},   \qquad  \sigma^{- +}_{0} = q^{-\frac12}
\eeq
in a weight basis. Then $(\sigma_i)^\a{}_\b$ are
\beqa
\sigma_{-1} = \left(\begin{array}{cc} 0 & q^{\frac12}\sqrt{[2]_q}\\ 
                            0 & 0 \end{array}\right),  \quad
\sigma_{0} = \left(\begin{array}{cc} -q^{-1} & 0 \\ 
                                                0 & q \end{array}\right), \quad
\sigma_{1} = \left(\begin{array}{cc} 0 & 0 \\ 
                           - q^{-\frac12}\sqrt{[2]_q} & 0 \end{array}\right).
\label{sigma-matrices}
\eeqa
They satisfy
\beqa
\sigma_i \sigma_j &=& - \ep_{ij}^k \; \sigma_k + g_{ij} \label{sigma-cliff}\\
\pi(u_1) \sigma_i \pi(S u_2) &=& \sigma_j \pi^j_i(u),
\eeqa
for $u \in U_q(su(2))$, where $\ep_{ij}^k$ is defined below, and
$\pi$ denotes the appropriate representation.

The invariant tensor $g^{ij}$ for the spin 1 representation 
satisfies by definition
\beq
\pi^i_k(u_1)\; \pi^j_l(u_2)\; g^{kl} = \ep(u) g^{ij}
\label{invar-g}
\eeq
for $u \in U_q(su(2))$. It is given by
\beq
g^{1 -1} = -q^{-1}, \;\; g^{0 0} = 1, \;\; g^{-1 1} = -q,
\label{g-explicit}
\eeq
all other components are zero.
Then $g_{ij} = g^{ij}$ satisfies  
$g^{ij} g_{jk} = \d^{i}_{k}$, and 
$g^{ij} g_{ij} = q^2 +1 + q^{-2} = [3]_q$.

The  Clebsch--Gordon coefficients
for $(3) \subset (3) \otimes(3)$, i.e. the $q$--deformed structure 
constants, are given by 
\beq
\begin{array}{ll} 
\ep^{1 0}_{1} =  q^{-1}, & \ep^{0 1}_{1} = -q,  \\
\ep^{0 0}_{0} = -(q-q^{-1}), & \ep^{1 -1}_{0} = 1 = -\ep^{-1 1}_{0}, \\
\ep^{0 -1}_{-1} = q^{-1}, & \ep^{-1 0}_{-1} = -q,
\end{array}
\label{C-ijk}
\eeq
and $\ep_{i j}^{k} := \ep^{i j}_{k}$. They have been normalized such that 
$\sum_{ij} \ep_{ij}^{n} \ep^{ij}_{m} = [2]_{q^2} \d^n_m$.
The $q$--deformed totally ($q$--)antisymmetric 
tensor is defined as follows: 
\beq
\ep^{ijk} = g^{in} \ep_n^{jk} = \ep^{ij}_n g^{nk}.
\label{q-epsilon}
\eeq
It is invariant under the action of $U_q(su(2))$, 
\beq
\pi^i_l(u_1)\; \pi^j_m(u_2)\; \pi^k_n(u_3)\; \ep^{lmn} = \ep(u) \ep^{ijk}.
\eeq

\subsection{Some proofs}\label{app:proof}

{\bf The form of ${\tilde \mathbf{\si}}_\mu$}
\vskip.3cm

Here we derive the form of ${\tilde \si}_\mu$.
Instead of a brute force calculation, we can argue as follows:
let 
\beq
\tilde \sigma_i:= -\RR_2 q^H S(\sigma_i) S(\RR_1),
\eeq
and observe that 
\beqa
u_1 \tilde \sigma_i S u_2 &=& -\RR_2 u_2 q^H S(\sigma_i) S(u_1) S(\RR_1) \non
  &=& -\RR_2 q^H S( u_1 \sigma_i S u_2) S(\RR_1) = 
     \tilde \sigma_j \pi^j_i(u).
\eeqa
This implies that $\tilde \sigma_i = c \sigma_i$ for a constant $c \in \mathbf{C}$.
The latter can be found e.g. by first showing 
\beqa
\tilde \sigma_i &=& S(\RR_1 \sigma_i q^{-H} S^{-1}(\RR_2)) \non
  &=& S(\RR_1 \trr \sigma_i \; \RR_a q^{-H} S^{-1}(\RR_2 \RR_b)) \non
  &=& S(\RR_1 \trr \sigma_i \;(\RR_a S \RR_b q^{-H}) S^{-1} \RR_2) \non
  &=& S(\RR^{-1}_1\trr \sigma_i \;\RR^{-1}_2) (S \RR_b \RR_a q^H) \non
  &=& S(\sigma_j {L^-}^j_i) v
\eeqa
where $v$ is the Drinfeld--Casimir 
\refeq{v}, and 
${L^-}^j_i = \pi^j_i(\RR^{-1}_1) \RR^{-1}_2$. 
On the spin 1/2 representation, $v$ takes the value $v=q^{-3/2}$. 
Using $\hat R^{ij}_{kl} g^{kl} = q^{-4} g^{ij}$ 
and $\Delta {L^-}^j_i = {L^-}^j_k \otimes {L^-}^k_i$, one can now verify
$g^{ij} \tilde \sigma_i\tilde \sigma_j = $ 
$q^4 v^2 g^{ij} \sigma_i\sigma_j$, which implies 
$c = q^2 v = q^{1/2}$. 

\vskip.5cm
{\bf Proof of (\ref{M-traf-l})}\label{app:cop}

\beqa 
&&(u^L\otimes u^R)\trr X\Xt{}= 
\cdot
((u^L_1 \otimes R_y u^R_a R^{-1}_j) \otimes (R_x u^L_2 R^{-1}_i \otimes
 u^R_b))\trr_\cF (X\otimes \Xt{})\non
&&=u^L_1 X S(R_y u^R_a R^{-1}_j) R_{\b} u^R_b R^{-1}_\d \Xt{}
S(R_\a R_x u^L_2 R^{-1}_iR^{-1}_\g)\non
&&= 
u^L_1 X S(R^{-1}_j) S(u^R_a){S(R_y) R_{\b}} u^R_b R^{-1}_\d \Xt{}
S(R^{-1}_\g) S(R^{-1}_i) S(u^L_2 ) {S(R_x) S(R_\a)}=\non
&&
= u^L_1 X S(R^{-1}_j) S(u^R_a) {R_y R_{\b}^{-1}} u^R_b R^{-1}_\d \Xt{}
S(R^{-1}_\g) S(R^{-1}_i) S(u^L_2 ) {R_x R_\a^{-1}}\non
&&= u^L_1 X S(R^{-1}_j) {S(u^R_a) u^R_b} R^{-1}_\d \Xt{}
S(R^{-1}_\g) S(R^{-1}_i) S(u^L_2 ) 
 \non
&&= u^L_1 X S(R^{-1}_j) R^{-1}_\d \Xt{}
S(R^{-1}_\g) S(R^{-1}_i) S(u^L_2 )
= u^L_1 X R_j R^{-1}_\d \Xt{}
S(R_i R^{-1}_\g)  S(u^L_2 ) \non
&&= u^L_1 X \Xt{}  S(u^L_2 )
\eeqa
 as desired.

\end{appendix}

\end{document}